\documentclass[reprint,aps,prb,twocolumn,groupedaddress,nobibnotes]{revtex4-1}
\usepackage{amssymb}
\usepackage{graphicx}
\usepackage[caption=false]{subfig} 
\usepackage{float}
\usepackage{amsmath}
\usepackage{dcolumn}
\usepackage{color}
\usepackage{graphicx}
\usepackage[pdftex,colorlinks=true,linkcolor=red,citecolor=blue]{hyperref}
\usepackage{soul}

\begin{document}

\title[]{Weak antilocalization effect due to topological surface states in Bi$_2$Se$_{2.1}$Te$_{0.9}$}
\author{K. Shrestha$^{1}$}
\email[Corresponding E-mail:]{keshav.shrestha@inl.gov}
\author{D. Graf$^{2}$}
\author{V. Marinova$^{3}$}
\author{B. Lorenz$^{4}$}
\author{C. W. Chu$^{4,5}$}
\affiliation{$^{1}$Idaho National Laboratory,2525 Fremont Ave, Idaho Falls, ID 83401, USA}
\affiliation{$^{2}$National High Magnetic Field Laboratory, Florida State University, Tallahassee, Florida 32306-4005, USA} 
\affiliation{$^{3}$Institute of Optical Materials and Technology, Bulgarian Academy of Sciences, Academy G. Bontchev Street 109, Sofia 1113, Bulgaria} 
\affiliation{$^{4}$TCSUH and Department of Physics, University of Houston, 3369 Cullen Boulevard, Houston, Texas 77204-5002, USA} 
\affiliation{$^{5}$Lawrence Berkeley National Laboratory, 1 Cyclotron Road, Berkeley, California 94720, USA} 

\begin{abstract}
  We have investigated the weak antilocalization (WAL) effect in $p$-type Bi$_2$Se$_{2.1}$Te$_{0.9}$ topological system. The magnetoconductance shows a cusp-like feature at low magnetic fields, indicating the presence of the WAL effect. The WAL curves measured at different tilt angles merge together when they are plotted as a function of the normal field components, showing that surface states dominate the magnetoconductance in the Bi$_2$Se$_{2.1}$Te$_{0.9}$ crystal. We have calculated magnetoconductance per conduction channel and applied the Hikami-Larkin-Nagaoka formula to determine the physical parameters that characterize the WAL effect. The number of conduction channel and phase coherence length do not change with temperature up to $T$ = 5 K. In addition, the sample shows a large positive magnetoresistance that reaches 1900\% under magnetic fields of 35 T at $T$ = 0.33 K with no sign of saturation. The magnetoresistance value decreases with both increasing temperature and tilt angle of the sample surface with respect to the magnetic field. The large magnetoresistance of topological insulators can be utilized in future technology such as sensors and memory devices.

\end{abstract}


\maketitle
\section*{Introduction}
\indent Three-dimensional topological insulators (TIs) have attracted significant attention as they possess topologically protected metallic states on their surfaces, known as surface states\cite{Hasan,Qi,Ando,Ando1,Cava}. These metallic surface states in 3D TIs originate as a result of the nontrivial topology of the bulk band structure and are protected by time-reversal symmetry. Due to the topological protection of these metallic states, surface-state electrons have very high mobility and are less sensitive to impurities (if the impurity is not magnetic). Recently, very large magnetoresistance and high mobility have also been observed in many topological systems\cite{Yan,Shrestha,Wang}, making TIs not only a playground for understanding novel quantum phenomena but promising candidates for future electronic devices as well. Many bismuth-based TIs have been theoretically predicted and have later been experimentally verified by surface-sensitive techniques such as angle-resolved photoelectron spectroscopy and scanning electron microscopy\cite{Xia,Chen,Hsieh}. Electrical transport (or magnetization) measurements under high magnetic fields have often been used to study topological systems. In the presence of magnetic fields, electrical conductivity (or magnetization) shows quantum oscillations known as Shubnikov de Haas (de Haas van Alphen) effects\cite{Kittel,Ashcroft,Shoenberg}. By analyzing oscillations at different tilt angles of the sample with respect to magnetic fields, one can map the two-dimensional Fermi surface of surface states or the three-dimensional Fermi surface of bulk states and thus study many additional properties\cite{Wang,Shrestha1}. However, transport studies of surface states in 3D TIs are always hindered due to the presence of the parallel bulk conduction channel that arises as a result of crystal defects and imperfections\cite{Qu,Analytis,Eto,Cao}. Several efforts have been made to grow purer topological crystals by modifying the crystal growth technique, compensating excess bulk carriers by doping Sb or Ca ions, and extending research from binary to ternary topological compounds \cite{Taskin, Gaku, Hor, Bao, Xu, Lin}. \\
\indent In our recent magnetotransport studies\cite{Shrestha2,Shrestha3} on metallic Bi$_2$Se$_{2.1}$Te$_{0.9}$ single crystals, we have observed well separated signals from the surface and bulk states. The surface states dominate at low fields (below 7 T) and bulk states at high fields. The crossover between these signals takes place at 14 T. These results have shown that it may be possible to characterize surface-state properties even if the bulk is metallic.\\
\indent Due to the presence of strong spin-orbit coupling in topological materials, their magnetoconductance often shows a weak antilocalization (WAL) effect\cite{He,Shrestha4}. As a quantum correction to a classical conductance, a WAL effect in topological insulators may originate due to spin-orbit coupling in either the surface or bulk states. Whether the WAL effect originates from surface or bulk states can be determined by measuring the magnetoconductance at different angles between the sample and the magnetic field direction. Recently, numerous topological systems have been successfully investigated by means of the WAL effect\cite{Taskin,Shekhar} where physical properties, such as the phase coherence length and the number of conduction channels, were determined; such measurements are not possible using the quantum oscillations method. It would therefore be interesting to extend the study of the Bi$_2$Se$_{2.1}$Te$_{0.9}$ sample by the WAL method.\\
\indent In this work, we have carried out magnetoresistance studies on a Bi$_2$Se$_{2.1}$Te$_{0.9}$ single crystal under high magnetic fields up to 35 T. The sample shows a large non-saturating magnetoresistance that reaches 1900\% under 35 T at $T$=0.33 K. Magnetoconductance in low magnetic fields shows a cusp due to the WAL effect. From the angle dependence of the WAL curves, we prove the presence  of topological surface states in a Bi$_2$Se$_{2.1}$Te$_{0.9}$ single crystal. We have estimated several physical parameters using the Hikami-Larkin-Nagaoka formula and have studied their temperature dependence as well.
\section*{Experimental}
High-quality Bi$_2$Se$_{2.1}$Te$_{0.9}$ single crystals were grown using the modified Bridgman method. Stoichiometric amounts of high purity Bi (99.9999\%), Se (99.9999\%), and Te (99.9999\%) were mixed together and enclosed in quartz ampoules. The mixture melts at 875 $^\circ$C and was kept at this temperature for 2 days. The molten mixture was slowly cooled to 670 $^\circ$C at a rate of 0.5 $^\circ$C/h and then to room temperature at a rate of 10 $^\circ$C/h. A shiny plate-like single crystal was selected from the boule of crystals. We used Scotch tape to peel out a very thin layer of the sample. Typical thickness of the sample is $\sim$0.05 mm. To ensure safe handling of this sample, it was attached to a magnesium oxide (MgO) substrate using GE varnish. Six gold contact pads were sputtered on the sample and platinum wires were attached to these gold contact pads using silver paint to carry out the standard longitudinal and Hall measurements.\\
\indent Transport measurements of the Bi$_2$Se$_{2.1}$Te$_{0.9}$ single crystal under magnetic fields up to 7 T were performed in a Physical Properties Measurement System (PPMS) at the Texas Center for Superconductivity at the University of Houston. The field range was extended to 35 T by performing measurements at the National High Magnetic Field Laboratory (NHMFL), Tallahassee, Florida. The angle-dependence measurements were carried out by mounting the sample on a rotating platform on a standard probe designed at NHMFL. Longitudinal and Hall resistances were measured using a lock-in technique in which a Keithley (6221) source meter provides AC current of amplitude 1 mA at a certain frequency, 47.77 Hz and the lock-in amplifier (SR-830) measures the voltage signal at the same frequency. The standard probe with the sample mounted on it was inserted into a $^3$He Oxford cryostat that sits into the bore of a resistive magnet with a maximum field of 35 T. A Hall sensor was used to calibrate the position of the sample with respect to the direction of the applied field.
\section*{Results and Discussion}
\begin{figure}
  \centering
  \includegraphics[width=1.0\linewidth]{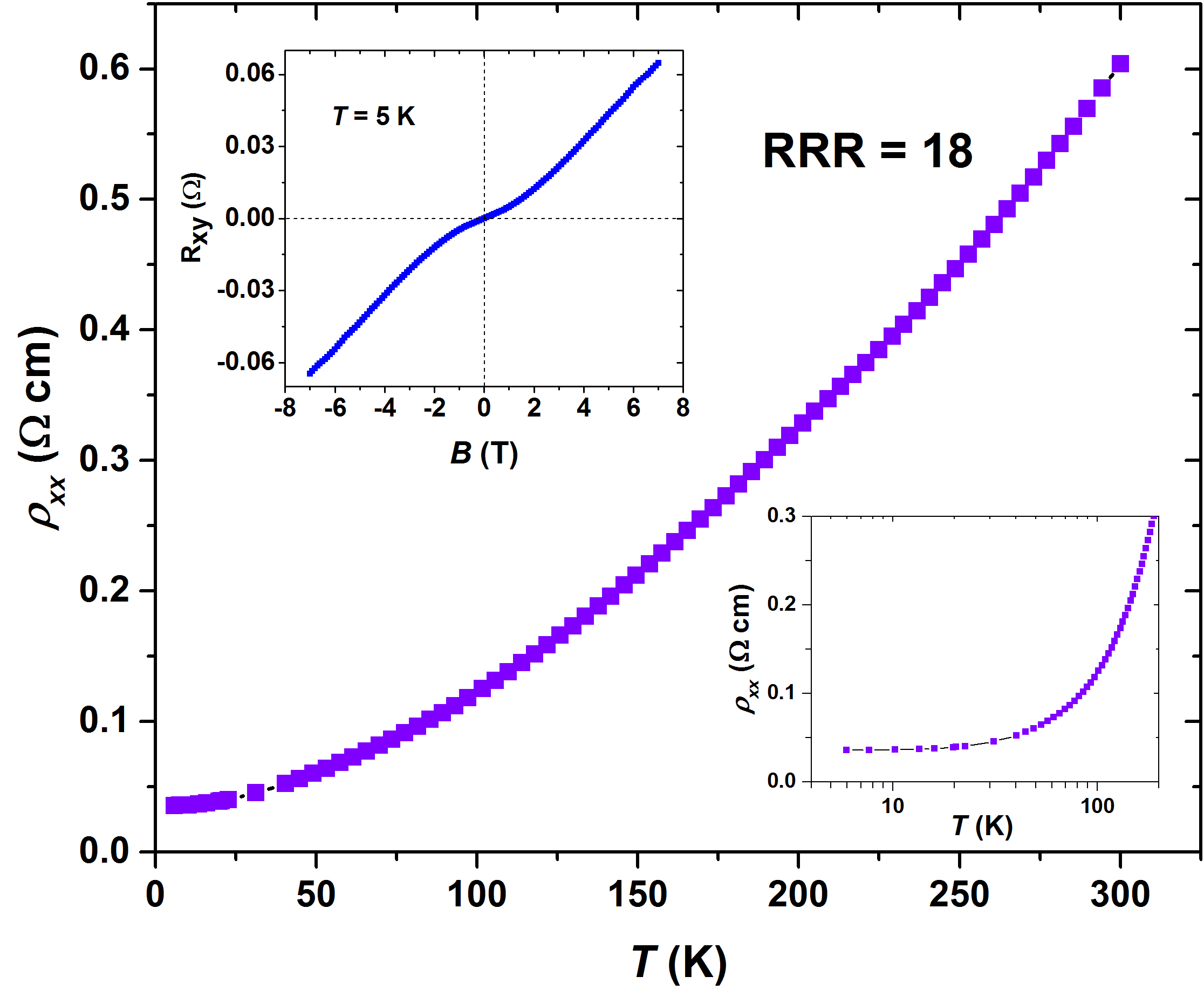}
  \caption{(Color online) Temperature dependence of longitudinal resistivity of a Bi$_2$Se$_{2.1}$Te$_{0.9}$ single crystal. Upper inset: Hall measurement data of Bi$_2$Se$_{2.1}$Te$_{0.9}$  at $T$=5 K. Lower inset: the resistivity in low temperature range with a logarithmic temperature axis.}\label{Fig1}
\end{figure}
\indent Figure [1] shows the temperature dependence of longitudinal resistivity of a Bi$_2$Se$_{2.1}$Te$_{0.9}$ single crystal. The sample shows a metallic behavior from 300 to 5 K. The high value of the residual resistance ratio, RRR=$\rho_{xx}$(300 K)/$\rho_{xx}$(5 K)=18 indicates good crystalline quality. This RRR value is comparable with those of Bi$_2$Se$_{2.1}$Te$_{0.9}$ single crystals used in our previous studies\cite{Shrestha2,Shrestha3}. The lower inset shows a zoomed in picture of the resistivity in low temperature range with the logarithmic temperature axis. The resistivity curve is almost flat below $T$ = 100 K. Such low temperature behavior of resistivity is also seen in other topological systems and considered as due to the dominance of topological surface states in this temperature regime\cite{Bansal, Chen, Chiatti}. The upper inset shows Hall measurements at $T$=5 K. The positive slope of the Hall resistance reveals the presence of hole-like bulk charge carriers. The Hall resistance shows the non-linear behavior near the origin, $B$=0, indicating the presence of multiband effects (hole and electron bands), as has been observed in other bismuth-based topological systems\cite{Qu,Shrestha2}. From the slope of the Hall data, we have estimated the bulk carrier concentration to be $\approx$ 8.7$\times$10$^{18}$cm$^{-3}$ at 5 K.\\
\begin{figure}
  \centering
  \includegraphics[width=1.0\linewidth]{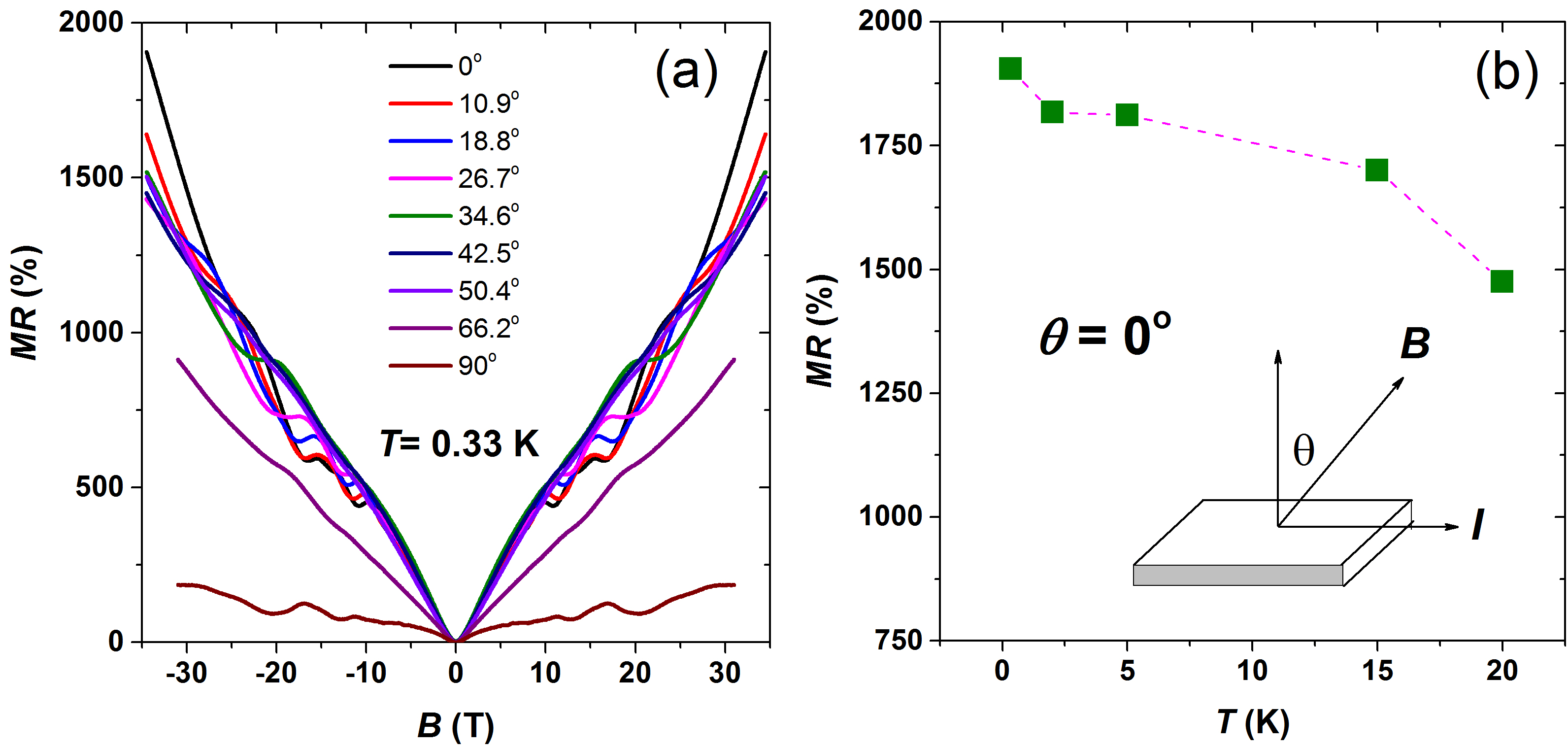}
  \caption{(Color online) (a) Angle dependence of MR of a Bi$_2$Se$_{2.1}$Te$_{0.9}$ single crystal expressed as a percentage in magnetic fields up to 34.5 T at $T$=0.33 K.(b) MR value at highest magnetic field at different temperatures measured at $\theta$=0$^{o}$. Inset: $\theta$ is the angle between magnetic field and normal to the sample surface.}\label{SdH}
\end{figure}
\indent The magnetoresistance of a Bi$_2$Se$_{2.1}$Te$_{0.9}$ single crystal was measured under high magnetic fields up to 35 T at NHMFL. Figure [2(a)] shows the magnetoresistance of Bi$_2$Se$_{2.1}$Te$_{0.9}$ measured at different tilt angles, $\theta$. Here, the angle $\theta$ is defined as the angle between the magnetic field direction and the perpendicular to the sample surface, as shown in the inset to Fig. [2(b)]. Magnetoresistance is expressed in percentage as MR=\big[${\rho}_{xx}(B)$/${\rho}_{xx}(0)$-1\big]$\times$100\%, where ${\rho}_{xx}(0)$ and ${\rho}_{xx}(B)$ are resistivity values at zero and $B$ applied field, respectively. The sample shows positive MR that increases linearly with magnetic field. MR reaches as high as 1900\% under 35 T at $\theta$=0$^{o}$ with no sign of saturation. At low magnetic field regime, MR shows a sharp cusp-like feature which indicates the presence of the weak antilocalization effect (WAL) in the Bi$_2$Se$_{2.1}$Te$_{0.9}$ sample. We will discuss the WAL effect later in detail. It should be noted that MR shows clear quantum oscillations in fields above 10 T. The oscillations have two frequencies at $F_1$$\approx$26 T and $F_2$$\approx$55 T in the frequency spectrum. From the angle dependence of $F_1$ and $F_2$ and Berry phase calculations, we have already resolved the origin of these frequencies in our previous studies\cite{Shrestha2,Shrestha3}. The MR value strongly depends on $\theta$. MR is maximum at $\theta$=0$^{o}$, and it decreases gradually at higher $\theta$ values. Similarly, MR decreases with increase in temperature, as shown in Fig. [2(b)]. At $T$=20 K, MR=1475\%, which is almost $\frac{2}{3}$ of the MR value at $T$=0.33 K. Topological materials are expected to show a large linear magnetoresistance due to the Dirac-like dispersion of surface states in the band structure\cite{Yan,Shrestha,Wang, Zhang}. Thus, an observation of large non-saturating magnetoresistance suggest the presence of topological surface states in Bi$_2$Se$_{2.1}$Te$_{0.9}$ sample.\\
\begin{figure}
  \centering
  \includegraphics[width=1.0\linewidth]{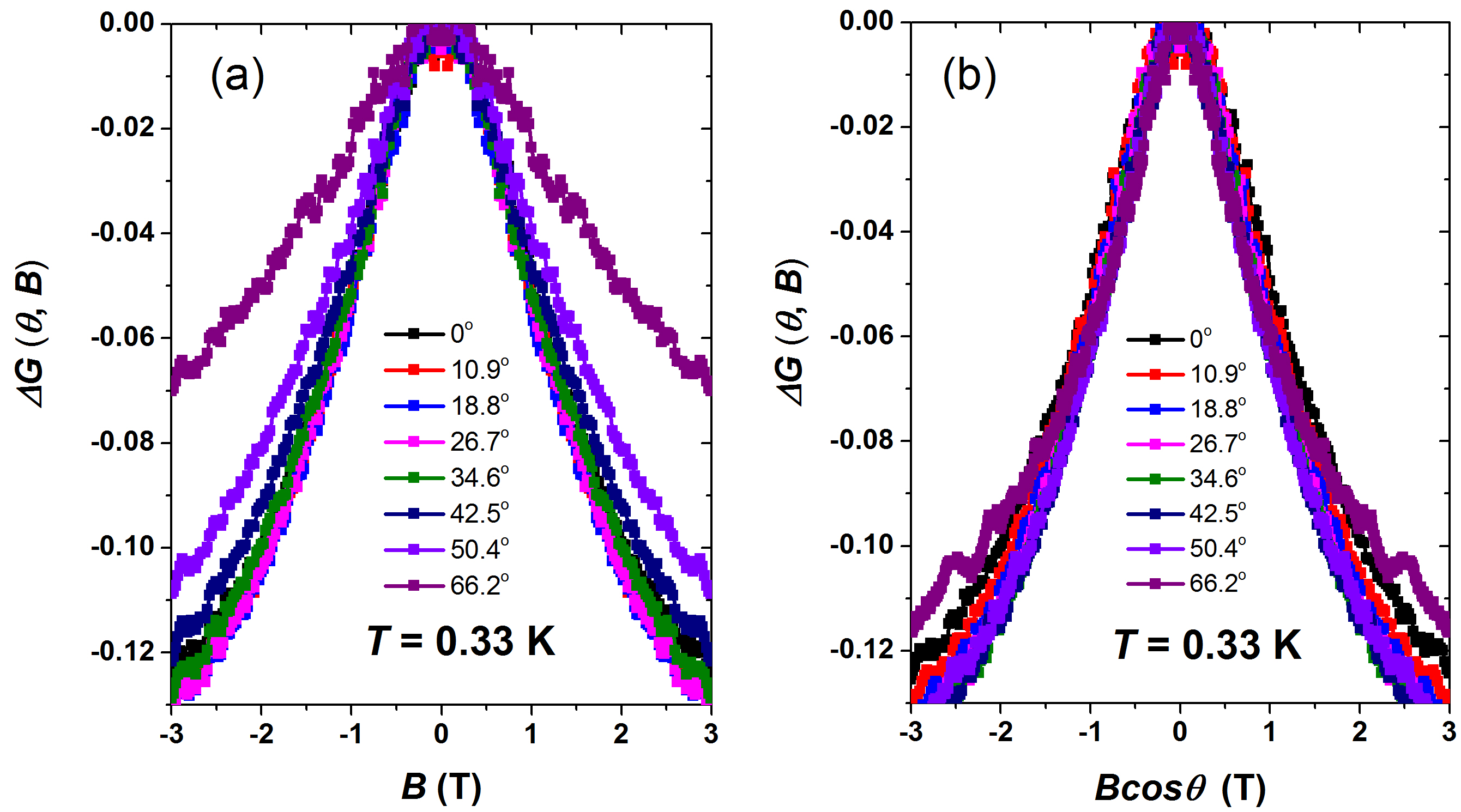}
  \caption{(Color online) Magnetoconductance curves of Bi$_2$Se$_{2.1}$Te$_{0.9}$ single crystal as a function of (a) $B$ and (b) $B$cos$\theta$ under fields up to 6 T at $T$=0.33 K.}\label{FFT}
\end{figure}
\indent In order to detect topological surface states in Bi$_2$Se$_{2.1}$Te$_{0.9}$, we have studied magnetoconductance at different tilt angles, $\theta$. The WAL-induced quantum corrections to magnetoconductance can be obtained as\cite{He}
\begin{equation}
\Delta G (\theta, B)=1/{\rho}_{xx}(\theta,B)-1/{\rho}_{xx}(\theta=90^{o},B),
\end{equation}\\
 where $\rho_{xx}(\theta,B)$ and $\rho_{xx}(\theta=90^{o},B)$ are resistivity values at a given $\theta$ value and at $\theta$=90$^{o}$, respectively. Figure [3(a)] shows $\Delta G (\theta, B)$ of the Bi$_2$Se$_{2.1}$Te$_{0.9}$ crystal measured along different tilt angles at $T$=0.33 K. Magnetoconductance shows a strong dependence on the $\theta$ values. All of the magnetoconductance curves merge together when they are plotted as a function of the normal component of magnetic fields, $Bcos\theta$, as shown in Fig. [3(b)]. This provides a strong evidence of the dominance of topological surface states in the magnetoconductance of the Bi$_2$Se$_{2.1}$Te$_{0.9}$ single crystal. This observation is consistent with the previous conclusions at low magnetic fields\cite{Shrestha2,Shrestha3}. \\
\indent For a deeper understanding of the WAL effect observed in Bi$_2$Se$_{2.1}$Te$_{0.9}$, we have used the Hikami-Larkin-Nagaoka (HLN) formula\cite{Hikami} and determined various physical parameters. According to the HLN formula, magnetoconductance can be described as,\\
\begin{equation}\label{Hikami}
\Delta G(\theta=0, B)=-\alpha\frac{e^2}{2\pi^{2}\hbar}\bigg[\Psi\bigg(\frac{1}{2}+\frac{\hbar}{4eL^{2}_{\phi}B}\bigg)-ln\bigg(\frac{\hbar}{4eL^{2}_{\phi}}\bigg)\bigg].
\end{equation}\\
\noindent Here $\Psi$ is the digamma function, and $L_{\phi}$ is the phase coherence length, which is the distance traveled by an electron before its phase is changed. The parameter $\alpha$=0.5 for a single coherent conduction channel. Equation (2) can be applied to the sample that shows a semiconducting-like behavior. However, in case of metallic-like samples we have to calculate magnetoconductance per conduction channels, i.e.  $\sigma=\Delta G/Z{^*}$ where $Z^*$ is the number of conduction layers\cite{Chiatti}. Following Chiatti $et$ $al.$\cite{Chiatti}, one 2D layer contributes a conductance value of $\sim$ $e^2/h$. One 2D layer is of about 2 quintuple layers with thickness of $\sim$2 nm. Thus, the number of conduction layers can be calculated as $Z^*$=$\times$ t/(2 nm) where $t$ is the sample thickness. Using equation (2) with our experimental data, the fitting parameters $L_{\phi}$ and $\alpha$ can be determined. Figure [4(a)] shows $\sigma$ vs $B$ data at $T$=0.33 K in a low field range (-1 to 1 T). The magnetoconductivity data is well described by the HLN formula as shown by the dashed curve. The fitting yields $L_{\phi}$=25.5 nm and $\alpha$=0.54 at $T$=0.33 K. Now, these values are comparable to those of previously reported data for other topological systems\cite{Xu1, Checkelsky, Chiu}. In order to explore the robustness of system in terms of conduction, we have calculated the parameters, $L_{\phi}$ and $\alpha$ at different temperatures as shown in Fig. [4(b, c)]. The parameter $\alpha$ and phase coherence length remain independent of temperature up to $T$=5 K.
\begin{figure}
  \centering
  \includegraphics[width=1.0\linewidth]{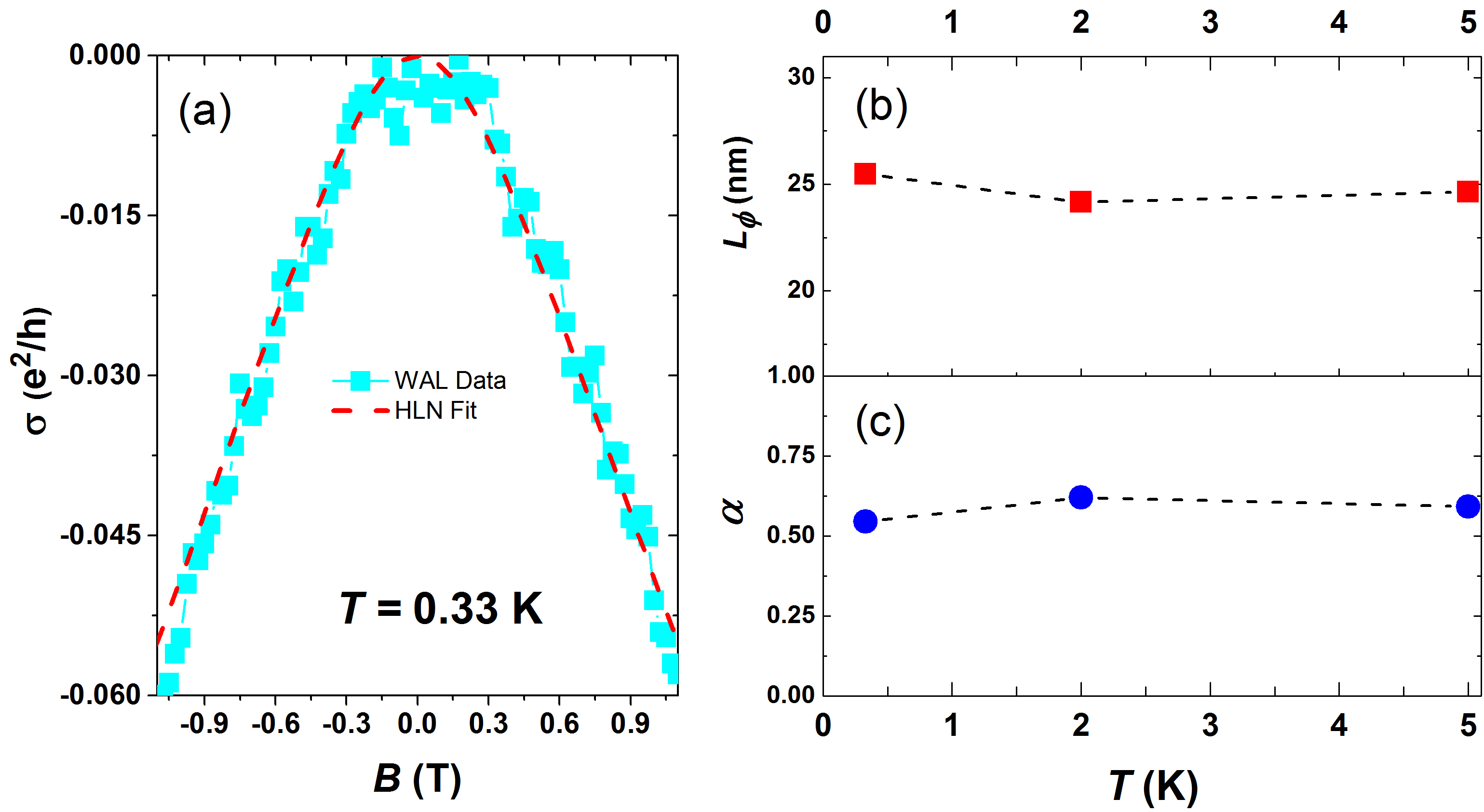}
  \caption{(Color online) (a) Magnetoconductance curve of Bi$_2$Se$_{2.1}$Te$_{0.9}$ single crystal within (-1 to 1 T) field range at $\theta$=0$^{o}$. The dashed curve shows the best fit curve obtained using the HLN formula. (b, c) Temperature dependence of $\alpha$ and phase coherence length $L_{\phi}$ respectively.}\label{Fig4}
\end{figure}
\section*{Conclusion}
\indent One of the biggest challenges in transport studies of bismuth-based topological systems is the bulk state conduction, which interferes with the surface conduction channel. Since the bulk states effect is larger than that of surface, it is challenging to detect the surface states signal and study their properties by transport measurements. In our previous works\cite{Shrestha2,Shrestha3}, we have separated the surface states signal by detailed quantum oscillations analyses. In this work, we have another transport measurement technique, the weak antilocalization (WAL) effect, to detect surface states in metallic Bi$_2$Se$_{2.1}$Te$_{0.9}$ single crystal.
The WAL curves at different tilt angles with respect to the magnetic field scale with the normal component of magnetic fields, further confirming the dominance of topological surface states in magnetoconductivity of Bi$_2$Se$_{2.1}$Te$_{0.9}$ sample. In order to investigate the WAL effect further, we have applied the Hikami-Larkin-Nagaoka formula to the magnetoconductivity data and determined various physical parameters. We have calculated $\alpha$=0.54 and phase coherence length $l_{\phi}$ $\sim$ 25 nm at $T$=0.33 K. The values of $\alpha$ and $l_{\phi}$ remain almost constant while increasing temperature up to $T$ = 5 K. In addition to the WAL effect, the Bi$_2$Se$_{2.1}$Te$_{0.9}$ sample shows a large positive magnetoresistance that reaches 1900\% under 35 T and $T$=0.33 K without any sign of saturation. Large magnetoresistance of Bi$_2$Se$_{2.1}$Te$_{0.9}$ makes it a suitable candidate for technological use in many future electronics such as sensors, spintronics and memory devices.\\

\section*{acknowledgements}
This work is supported in part by the U.S. Air Force Office
of Scientific Research Grant FA9550-15-1-0236, the T. L. L. Temple Foundation, the John J. and Rebecca Moores Endowment, and the State of Texas through the Texas Center for Superconductivity at the University of Houston. V. Marinova acknowledges support from the Bulgarian Science Fund project DN 08/9. A portion of this work was performed at the National High Magnetic Field Laboratory, which is supported by National Science Foundation Cooperative Agreement No. DMR-1157490 and the State of Florida. The work at Idaho National Laboratory is supported by Department of Energy, Office of Basic Energy Sciences, Materials Sciences, and Engineering Division.

\bibliography{STS}

\end{document}